\documentstyle[A4,12pt,epsf]{article}
\begin{document}

\begin{flushright}
{\large PSI-PR-96-13}
\end{flushright}
\vspace*{2.0cm}

\begin{center}
{\Large\bf The {\boldmath $\pi\pi$} Final State Interaction 
in {\boldmath $K\to\pi\pi$}, {\boldmath $pp\to pp\pi\pi$} 
and Related Processes}\\
\vspace*{1.0cm}
        {\sc M.~P.~Locher, V.~E.~Markushin and H.~Q.~Zheng}\\
\vspace*{0.5cm}
        {\it Paul Scherrer Institute, 5232 Villigen PSI, Switzerland} \\ 
\vspace*{0.5cm}
April 10, 1996
\end{center}

\begin{abstract}
Final state interactions in the $S$--wave $\pi\pi$ system (I=0,2) are
re-examined on the basis of the  Omn\`es-Mus\-khe\-li\-shvili equation and
the coupled channel formalism.  The contributions to the pion scalar form
factor from $\rho$ and  $f_2(1270)$ exchange in the $t$--channel and from
the $f_0(980)$ $s$--channel resonance are separately evaluated and the role
of the nontrivial polynomial in the Omn\`es function in a coupled channel
situation is elucidated.  Applications are made to $K\to \pi\pi$ and $pp\to
pp\pi\pi$. It is found that the contribution from the $f_0$ resonance to the
form-factor is strongly reduced by a nearby zero.
\end{abstract}

\noindent

\section{Introduction}

  Final state interaction (FSI) in the $\pi\pi$ system plays an important
role  for many production reactions and meson decays.  
A case of long-standing  interest is the $\Delta I = 1/2$ rule in
$K\to\pi\pi$ decays. The experimental ratio of the decay amplitudes $A_I$
with isospin $I=0,2$ is \cite{PDT}
\begin{equation}
     \frac{A_0(K\to\pi\pi)}{A_2(K\to\pi\pi)} = 22 
\end{equation}
The calculated ratio is smaller \cite{Kpipi} by at least a factor of 3 where
this result includes perturbative QCD  and soft-gluon corrections at the
weak interaction vertex but no long-distance $\pi\pi$ FSI.  In this paper we
shall discuss the pionic FSI in the $S$-wave aiming at a concrete
application to the $\Delta I = 1/2$ rule for the  $K\to\pi\pi$ decay and the
pion production reaction $pp\to\pi\pi pp$. Our analysis shows general
features of FSI's which are relevant to other reactions involving pions or
other hadrons.

Several methods for the evaluation of FSI have been used in the literature.
In one approach rescattering diagrams are evaluated directly. At low
energies this has been done by applying chiral perturbation theory (CHPT)
\cite{GL85,T88}. The relevant application in our context is the calculation
of the scalar form-factor of the pion in next to leading chiral order at low
energies \cite{DGL90,GM91}. To extend the calculations to $s\sim 1$~GeV$^2$
$s$-channel  resonances and the coupling to the $K\bar{K}$ channel must
be included.
As a general tool the dispersive method  based on the
Omn\`es-Mus\-khe\-li\-shvili (OM) equation  \cite{M53,O58} has turned out to
be very efficient. It exploits analyticity and unitarity in order to connect
the production or decay amplitude (or its form-factor)  with the amplitude
of elastic $\pi\pi$ scattering. To solve the OM equation we shall take the
scattering phases either from phase shift analysis or from a theoretical
model. We shall choose a model which satisfies the requirements of unitarity
and analyticity, and hence the OM equation automatically. The  model with
parameters fitted to the experimental constraints is described in Sec.~3. 
  
  For the $K\to\pi\pi$ decay it was realized a long time ago that the
non-perturbative long-distance effects must be included. An enhancement of
about a factor of 2 in the $I=0$ amplitude was estimated to result from the
broad $\sigma(J^{PC}=0^{++})$ meson \cite{VZS}.   For $K\to\pi\pi$ the
attraction in the $I=0$ channel must be combined with the repulsion in the
$I=2$ channel which favours the $\Delta I=1/2$ rule. 
  The analysis was done in CHPT to one loop in \cite{GM91,BBEL}. 
Rescattering in simple potential model was evaluated in \cite{BJS,IMW90} 
without regard to the energy dependence of the form-factor. 
  An extensive study of the FSI effects in the $S$-wave $\pi\pi$ system in
production reactions and $J/\psi$ and $\psi '$ decays was conducted in
\cite{MP75,MP84,AMP87,MP93}.  Unitarity and analyticity of the production
amplitudes was taken into account in a self-consistent way.  It was noticed,
in particular, that a narrow resonance ($f_0$ in the present  notation)  in
the  $\pi\pi$ scattering phase $\delta^{I=0}_{J=0}(s)$  corresponds to a
shoulder in the $\pi\pi$ effective mass distribution in the reaction $pp\to
pp\pi\pi$ \cite{MP84,AMP87}.  The occurrence of a shoulder   rather than a
peak results from an interplay of the resonant  pole and a nearby zero. We
shall discuss this feature in detail in Sec.~3.
  Resonance phenomena in the $\pi\pi$ S-wave were emphasized  in \cite{RW88}
where the $f_0$ resonance was discussed within a single-resonance model for
the decay of a light higgs boson.  The prediction of a drastic enhancement
due to the $f_0$ resonance is in striking contrast with the findings for
the $pp\to pp\pi\pi$ reaction in \cite{AMP87}.
An analysis of the $\pi\pi$ final state interaction in the
framework of the coupled channel OM equation was performed in \cite{DGL90} 
for the decay  of a light higgs boson decay $H\to\pi\pi$.  
In this evaluation the $f_0$ resonance also produced significant effects 
far below the $K\bar{K}$ threshold.

  The dynamics of the $I=0$ $S$-wave $\pi\pi$ interaction is characterized by 
several overlapping resonances \cite{AMP87,MP93,To95}, narrow and broad. In
the present paper we shall analyze the relative importance of the dynamical
mechanisms in $\pi\pi$ scattering for the calculation of the form-factors
occurring in meson decays and in the pion pair production in $pp$
scattering.
  In Sec.~2 we prepare the ground with an evaluation of the OM equation for a
restricted energy range (the cut-off used excludes the $f_0$ resonance).
With respect to the pion dynamics we shall mainly use the picture of
\cite{ZB94} which combines the $\rho$ and $f_2$ exchanges in the $t$-channel
with the $f_0$ resonance in the $s$-channel. The phases of the $I=0,2$
$S$-wave scattering are reproduced quite accurately in this model. To
understand the role of the $f_0$ resonance for the calculation of the
form-factor in the $I=0$ channel we shall introduce a coupled channel ansatz
in Sec.~3. 
  The final state interaction effects in the $K\to\pi\pi$ decay are evaluated 
and the conclusions are presented in Sec.~\ref{Sec_Kpipi}.

\section{Form-factors from the Omn\`es -- Muskhelishvili equation}
\label{OM}

  The OM equation \cite{M53,O58} connects the form-factor $F(s)$ with 
the elastic final state scattering phase $\delta(s)$. For a single channel
problem  the OM equation is   
\begin{eqnarray}
  F^{-1}(s) & = & 1+ \frac{s}{\pi} \int_{4m_{\pi}^2}^{\infty} 
                  \frac{\delta(s') F^{-1}(s')}{s'(s-s')} ds' 
\label{EqOM}
\end{eqnarray}
where a once-subtracted form has been used.
The general solution of (\ref{EqOM}) has the form 
\begin{eqnarray}
  F(s) & = & P(s) \exp \left( \frac{s}{\pi}
             \int_{4m_{\pi}^2}^{\infty} \frac{\delta(s')}{s'(s'-s)}
             ds' \right) 
\label{FOM}
\end{eqnarray}
as long as 
\begin{equation}
\delta(s)\to \mbox{\rm const, \ \ \ }
{|F(s)|\over s}\to 0\  \mbox{\rm \ \ \ for \ } s\to \infty .   
\end{equation}
The polynomial  $P(s)$ is real for $s$ real. 
In special cases, like potential scattering without bound 
states, $P(s)$ is a constant, but in general additional 
information is required to determine it.  

\begin{figure}[htb]
\begin{center}
\mbox{\epsfxsize=60mm \epsffile{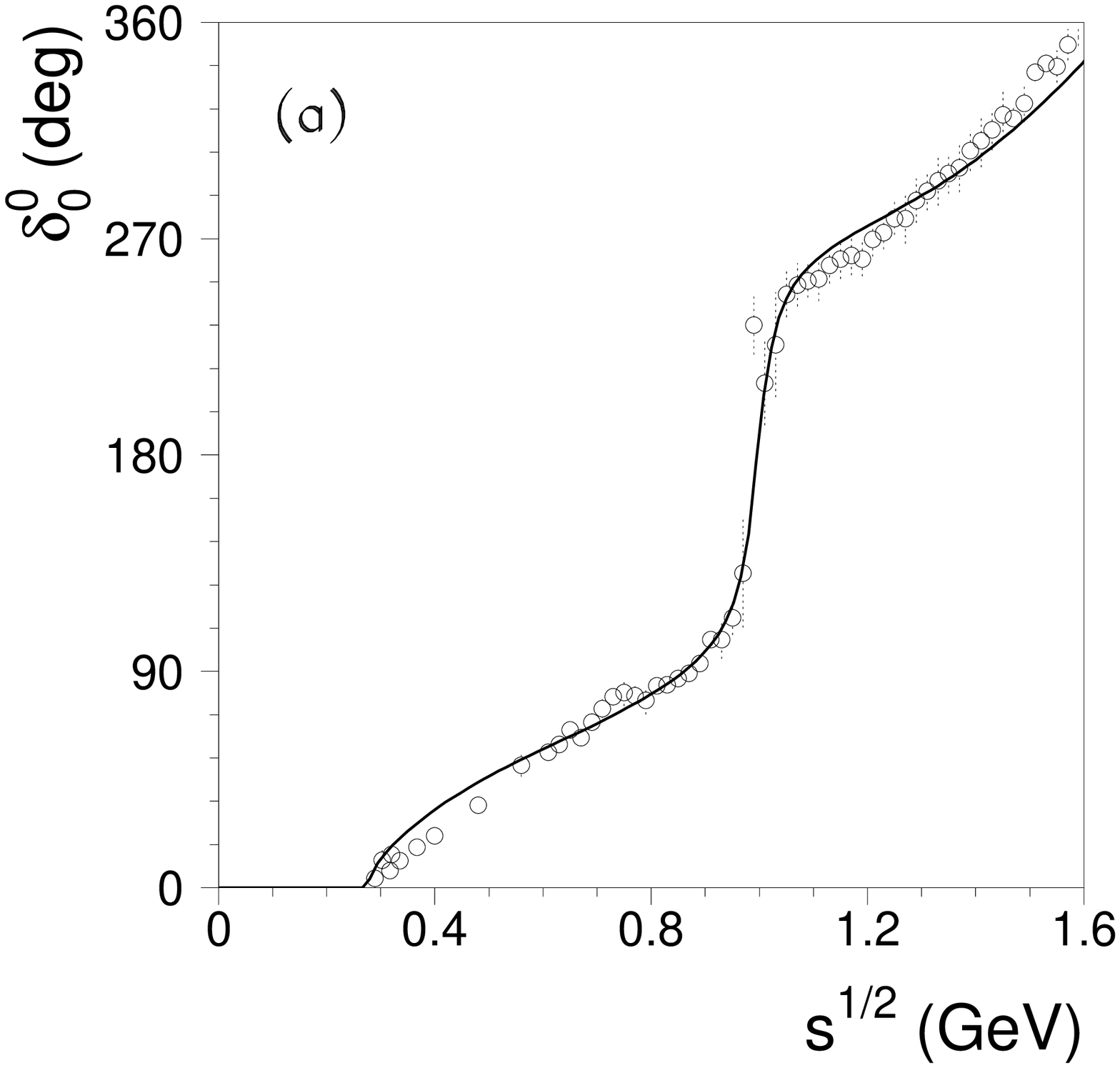}}
\hspace*{10mm}
\mbox{\epsfxsize=60mm \epsffile{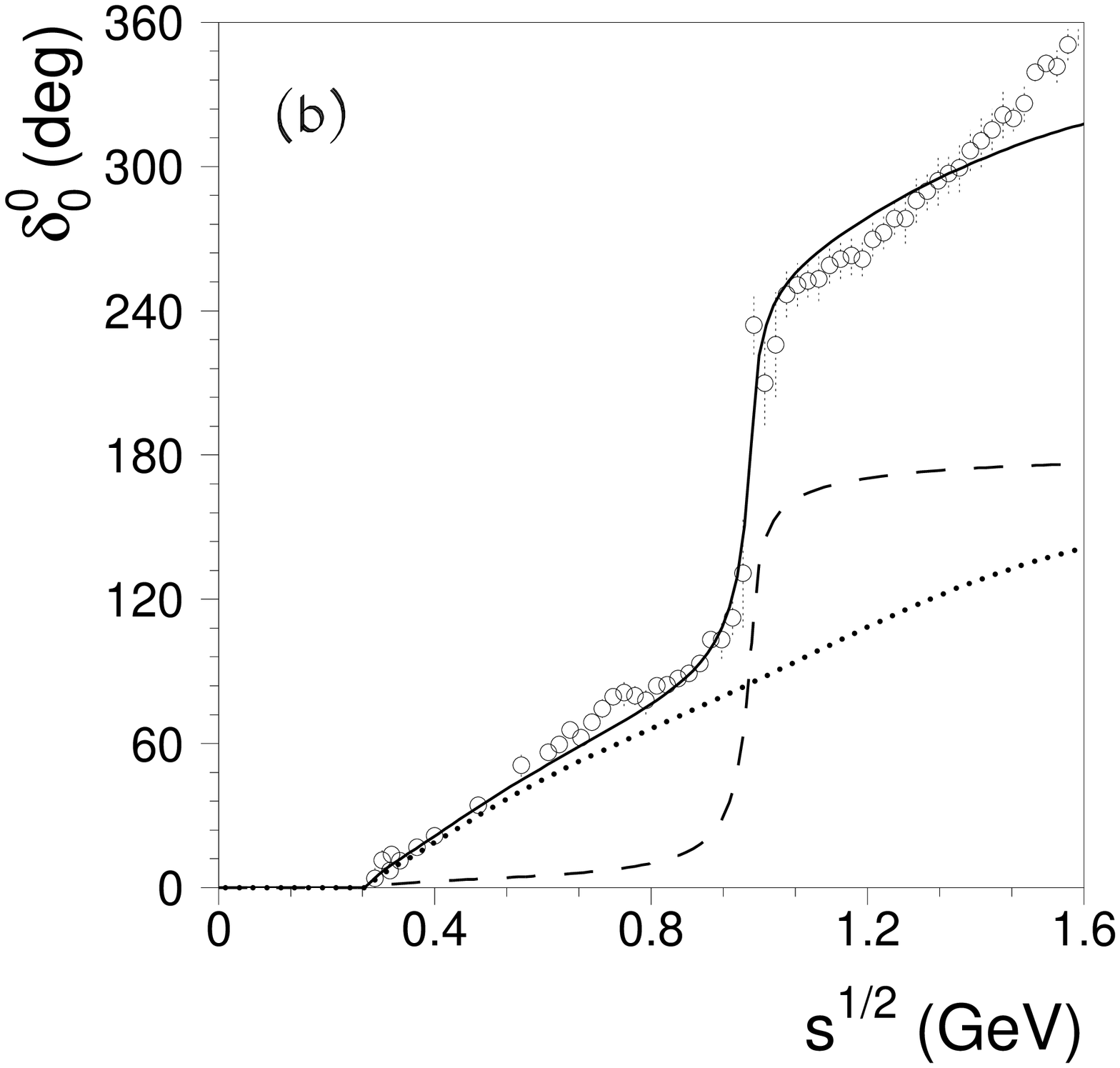}}
\vspace*{-10mm}
\caption{\label{PhaseFit}%
The $\pi\pi$ $S$-wave scattering phase $\delta^0_0$ 
vs. $\protect\sqrt{s}$:
(a) the K1 fit from  \protect\cite{AMP87},
(b) the meson exchange model described in the text
(solid line: the total phase; dashed line:
$\rho+f_2$ $t$-exchange; dotted line: $f_0$ resonance). 
The experimental data are from
\protect\cite{Gr74,Oc74,Ro77}.}
\end{center}
\end{figure}

  For $K\to\pi\pi$  in the simplest evaluations single channel $\pi\pi$
scattering data are used below  the $K\bar{K}$ threshold (the coupling to
the $4\pi$ channel is known to be small). Fig.~\ref{PhaseFit}a shows the
$\pi\pi$ $J=I=0$ scattering phase  $\delta^0_0(s)$ from the phase shift
analysis \cite{AMP87}.
  In Fig.~\ref{PhaseFit}b we show the same phase from the meson exchange model
mentioned earlier and developed in   \cite{ZB94,S60,BL71}. We briefly
recapitulate the ingredients for the benefit of the later discussion. The
phases in Fig.~\ref{PhaseFit}b correspond to unitarized $\rho$ and $f_2$
exchange with the $f_0$ resonance added. The individual contributions are
shown in the figure as explained in the capture. The Born term for the
$\rho$-exchange is 
\begin{eqnarray}
   T(s,t)^{I=0}_{BA} & = &
   2G \left( \frac{s-u}{m_\rho^2-t} + \frac{s-t}{m_\rho^2-u} \right) \\
   T(s,t)^{I=2}_{BA} & = & - \frac{1}{2} T(s,t)^{I=0}_{BA}
\label{TrhoBA}
\end{eqnarray}
where $m_{\rho}$ is the mass of the  $\rho$ meson,
$G=g_{\rho\pi\pi}^2/32\pi$, and $g_{\rho\pi\pi}$ is the $\rho\pi\pi$
coupling constant. The $I=2$ amplitude will be needed later. 
The $S$-wave projection is
\begin{equation}
  T_{BA-S}^{I=0}(s) = 4G \left[ \frac{2s + m_\rho^2 - 4m_\pi^2}{s - 4m_\pi^2}
   \ln \left( 1 + \frac{s-4m_\pi^2}{ m_\rho^2} \right)
   - 1 \right] 
\label{TS0}
\end{equation}
$K$-matrix unitarization is introduced by
\begin{eqnarray}
   T_{S}^{I}(s) & = & \frac{K_{S}^{I}(s)}{1-i \rho(s) K_{S}^{I}(s)}
\label{Trho}
\end{eqnarray}
where
\begin{eqnarray}
   K_{S}^{I}(s) & = & T_{BA-S}^{I}(s)
\label{Krho}
\end{eqnarray}
and 
$\rho(s)=(1-4m_{\pi}^2/s)^{1/2}$.
The coupling constant $g_{\rho\pi\pi}$ is determined from the
$\rho$ meson decay width in the crossed $I=1$ channel after 
$K$-matrix unitarization \cite{ZB94}. The corresponding value is 
$g_{\rho\pi\pi}=6.04$ which is close to the result obtained from the
KSFR relation \cite{KSFR},
$g_{\rho\pi\pi} = m_\rho/{\sqrt{2} f_\pi}$.

The corresponding expression for $f_2$ exchange in Born approximation 
\cite{ZB94} is 
\begin{eqnarray}
 K_{f_2}(s) & = & 2G_{f_2} \left\{ 
{-11\over 3}s - {2\over 3}m^2_{f_2}+4m_\pi^2 +  \right. \\ & & \left.
+\frac{(2s+m_{f_2}^2-4m_\pi^2)^2-(m_{f_2}^2-4m_\pi^2)^2/3}{s-4m_\pi^2}
\ln (1+{s-4m_\pi^2\over m_{f_2}^2})  \right\} 
\end{eqnarray}
where $G_{f_2}\simeq 0.19$~GeV$^{-2}$.

The $f_0$ resonance is included using the Dalitz-Tuan representation
\cite{DT60}, i.e. the  S-matrix is considered to be the  product
of the S-matrices corresponding to the individual mechanisms.   
The corresponding Breit-Wigner parametrization is taken from \cite{ZB94}:
\begin{eqnarray}
\label{resf}
    S(s) & = & 
    \frac{s - M_r^2 - i g_1 \rho_1(s) + i g_2 \rho_2(s)}{s 
             - M_r^2 + i g_1 \rho_1(s) + i g_2 \rho_2(s)}
\end{eqnarray}
where $\rho_1(s) = \sqrt{1-4m_{\pi}^2/s}$, 
$\rho_2(s) = \sqrt{1-4m_{K}^2/s}$, 
and the resonance parameters are  
$M_r=0.9535$~GeV, $g_1=0.1108$~GeV$^2$, $g_2=0.4229$~GeV$^2$, 
respectively. 
The scattering phase in the meson exchange model 
is a good description 
of  the data for
$s<1.4\;$GeV$^2$, see Fig.~\ref{PhaseFit}b.

\begin{figure}[htb]
\begin{center}
\mbox{\epsfxsize=60mm \epsffile{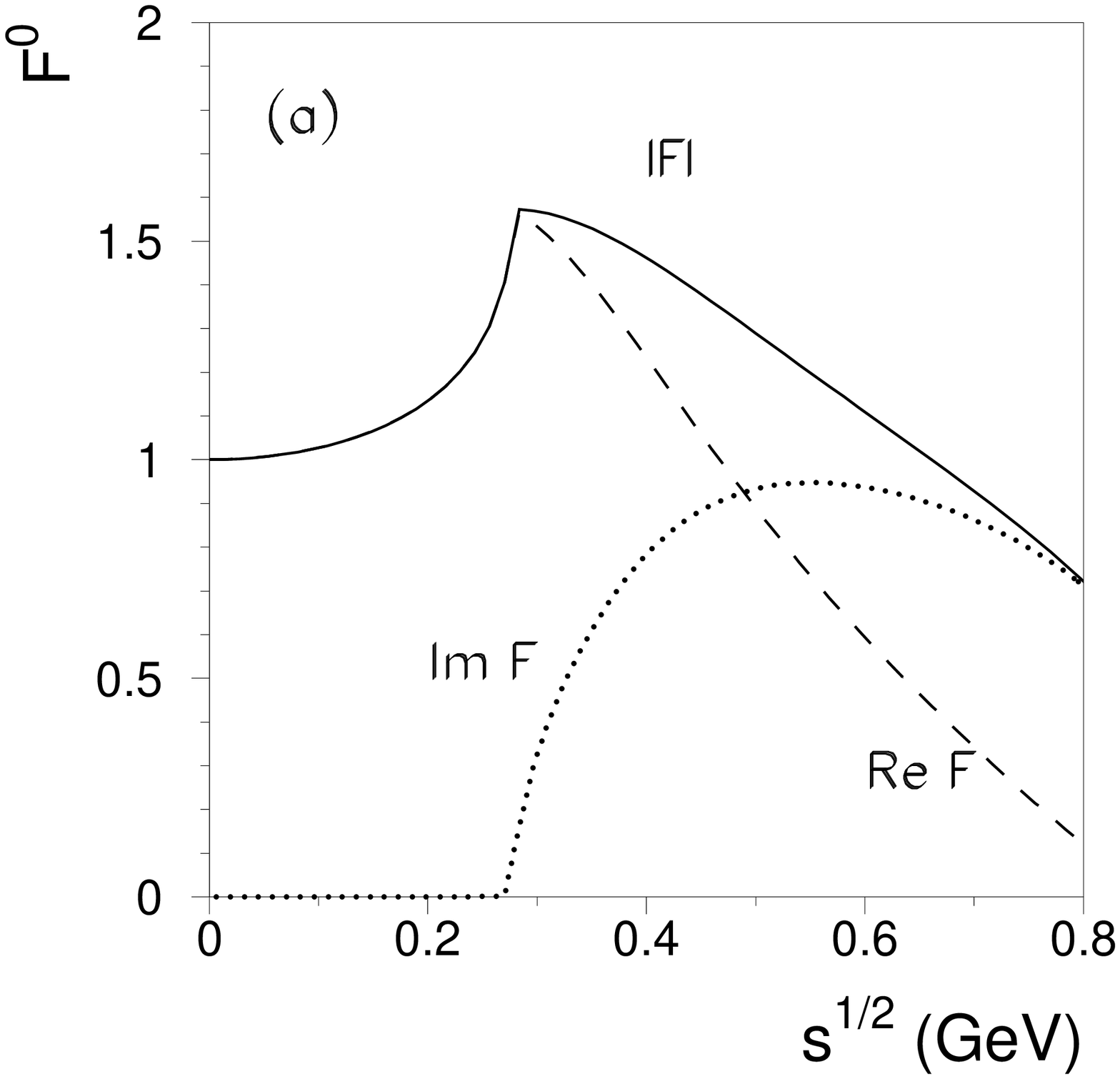}}
\hspace*{10mm}
\mbox{\epsfxsize=60mm \epsffile{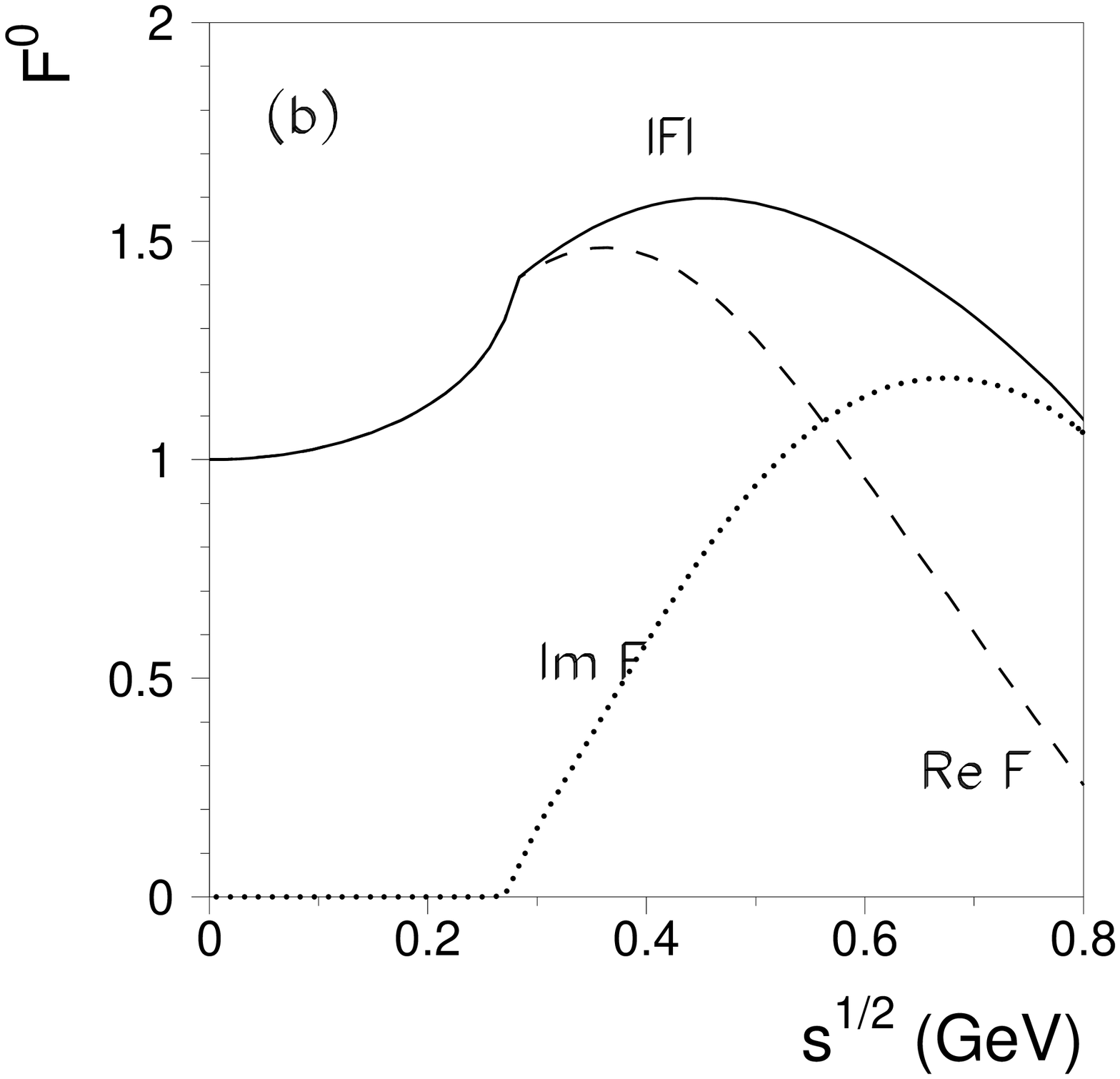}}
\vspace*{-10mm}
\caption{\label{FormFactorOML}
The pion scalar form-factor $F^{I=0}(s)$ vs. $\protect\sqrt{s}$
evaluated using the OM equation with the cut-off $\Lambda=0.975\;$Gev and
the $\pi\pi$ scattering phases 
as shown in Fig.~\protect\ref{PhaseFit}a and 
\protect\ref{PhaseFit}b, correspondingly.}
\end{center}
\end{figure}

  In order to show the sensitivity of the form-factors to the  variations in
the phase we shall evaluate the OM equation using the two sets  shown in
Fig.~\ref{PhaseFit}a and \ref{PhaseFit}b.  For small $s$ the integral of
Eq.~(\ref{FOM}) is dominated by low energies, see \cite{GM91}. As a first
step we evaluate in this section the OM equation with a cut-off $\Lambda$.
Choosing $\Lambda=0.975\;$GeV we exclude the $f_0$ and the $K\bar{K}$
threshold region as in some early applications.
We therefore write 
\begin{eqnarray}
F(s) & = & F_{\Lambda}(s) P(s)
\end{eqnarray}
where 
\begin{eqnarray}
F_{\Lambda}(s) & = & \exp \left(\frac{s}{\pi}
\int_{4m_{\pi}^2}^{\Lambda^2} \frac{\delta(s')}{s'(s'-s)} ds'
\right)\ .
\label{FOML}
\end{eqnarray}
The polynomial $P(s)$ represents the contribution from high energies and any
other dynamics not included so far. We observe that any additive
contributions in the phase lead to a multiplicative factor in $F$, see
eq.~(\ref{FOM}). We shall use 
\begin{eqnarray}
P(s)  = 1 + b s  \label{B}
\end{eqnarray}
where the  parameter $b$  is related to the scalar radius of the 
pion by
\begin{eqnarray}
F(s)  = 1 + \frac{1}{6} \langle r_s^2\rangle \, s \, .  
\label{Fs0}
\end{eqnarray}
  
When  plotting Fig.~\ref{FormFactorOML} we have adjusted the polynomial 
(\ref{B}) in order to have $\langle r_s^2\rangle=0.6$~fm$^2$ \cite{GM91} 
in both cases. This leads to 
$b=0.32\;$GeV$^{-2}$ for Fig.~\ref{FormFactorOML}a 
and $b=0.83\;$GeV$^{-2}$ for Fig.~\ref{FormFactorOML}b.
Figure ~\ref{FormFactorOML} shows that the form-factor is rather sensitive
to  the $\pi\pi$ scattering phase shift, one of the important parameters
being the scattering  length $a_0^0$. For the meson exchange model
$a_0^0=0.24 m_{\pi}^{-1}$ and the absolute  value of the form-factor
continues to rise between the $\pi\pi$ threshold and  $\sqrt{s}\approx
0.5\;$GeV. On the other hand the form-factor corresponding to  the phase in
Fig.~\ref{FormFactorOML}a displays a more prominent cusp due  to a larger
value of $a_0^0=0.51 m_{\pi}^{-1}$ and decreases above the  $\pi\pi$
threshold. 
For $\sqrt{s}<0.5$~GeV the result in Fig.~\ref{FormFactorOML}b
is close to the solution of the coupled-channel OM 
equation in \cite{GM91}.

\section{Protective Zero and the $f_0$ Resonance}
\label{ResZero}

\begin{figure}[htb]
\begin{center}
\mbox{\epsfxsize=60mm \epsffile{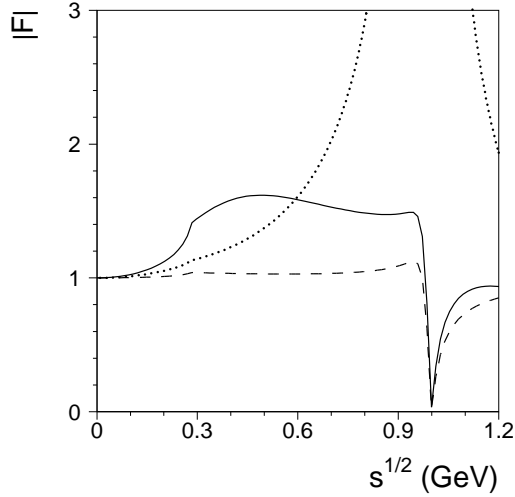}}
\vspace*{-10mm}
\caption{\label{FigFFrhof0}
The pion scalar form-factor $F^{I=0}(s)$ vs. $\protect\sqrt{s}$. 
The solid line corresponds to $\rho$ and $f_2$
exchange and an $f_0$ resonance including the polynomial (protective zero).
The $f_0$ resonance alone leads to the dotted line 
(OM equation without a polynomial) and to the dashed line with 
polynomial.}
\end{center}
\end{figure}

  For energies around $s=1\;$GeV$^2$ the truncation in the calculation of
the form-factor must be abandoned and the role of the $f_0$ and the
$K\bar{K}$ threshold discussed. 
The resonant part of the phase will be defined by 

\begin{eqnarray}
    \delta_{res}(s) & = & \mbox{\rm Arctan} \frac{g k(s)}{(M_r^2 - s)} 
\end{eqnarray}
corresponding to the resonance amplitude 
\begin{eqnarray}
    T_{res}(s) & = & \frac{g k(s)}{s - M_r^2 + i g k(s)}
\end{eqnarray}
where $k(s) = \sqrt{s-4m_{\pi}^2}/2$, leads to 
\begin{eqnarray}
  F_{res}(s) & = & \exp \left( \frac{s}{\pi}
             \int_{s_0}^{\infty}
             \frac{\delta_{res}(s')}{s'(s'-s)} ds' \right) = \\
    & = & \frac{M_r^2 + g m_{\pi}}{M_r^2 - s - i g k(s)} 
\label{naive}
\end{eqnarray}
Inserting this phase naively into Eq.~(\ref{FOML})
has the undesirable feature that
\begin{eqnarray}
 |F_{res}(s)| & \stackrel{s\to\pm\infty}{\to} & 0
\label{Fnaive}
\end{eqnarray}
rather than unity which would be expected  at high energies where the
resonance contribution should vanish.  The wrong asymptotic form is actually
imposed on the whole solution by means of phase additivity. In
Fig.~\ref{FigFFrhof0} the dotted line corresponds to the naive evaluation of
$F(s)$, with $P(s)$ being set to unity. Apart from the wrong asymptotics  it
is also seen that the $f_0$ resonance dominates the form-factor far outside
the resonance region $M_{f_0}\pm\Gamma_{f_0}$. Recall that the experimental
width is $\Gamma_{f_0}\approx 60\;$MeV. It is clear that this defect should
be compensated by a non-trivial polynomial  $P(s)$ in the solution of the OM
equation. 

\subsection{The OM equation for a resonance in
the Weisskopf-Wigner model}
\label{WWM}

  We shall study the modification required for a sensible  inclusion of a
direct channel resonance into the OM equation by means of a  very simple
coupled channel model. The following nonrelativistic ansatz, which is a
variant of the Weisskopf-Wigner (WW) model,  already has all the necessary
ingredients. Only one scattering channel is introduced which has no diagonal
potential. It will be denoted by its relative momentum $|k\rangle$. The only
interaction in the model results from the coupling to a bound state
$|b\rangle$ (effectively representing a second channel). We assume a channel
coupling of the form 
\begin{equation}
  \langle k | V | b \rangle = \gamma \xi(k) = \frac{\gamma}{k^2+\mu^2}
\end{equation}
where $\gamma$ is the coupling constant (dimension $[\gamma]=[k]^{3/2}$)
and $\mu$ characterizes the range of interaction.
The T-matrix satisfies the Lippmann-Schwinger equation
\begin{equation}
    T(E) = V \frac{|b \rangle \langle b|}{(E-E_r)}V (1 + G_0(E) T(E))
\end{equation}
where $G_0(E)$ is the free Green function and $E=k^2/2m$ ($m$ 
is the reduced mass). 
The solution for the scattering amplitude has the form
\begin{eqnarray}
  f(k) & = & - 2 m \langle k |T(E)| k \rangle =  \\ & = & 
       \frac{- 2 m \gamma^2 \xi^2(k)}{\displaystyle \frac{k^2}{2m} 
       - E_b - \gamma^2 D(k)}
\end{eqnarray}
with
\begin{eqnarray}
  D(k) & = & \langle b| V G_0(E) V | b \rangle
         = \frac{m}{\mu(k+i\mu)^2} \ .
\end{eqnarray}

In our model the form-factor $F(k)$ describing the final state interaction
is equal to the scattering wave function at zero distance according
to standard results from scattering theory \cite{Taylor}: 
\begin{eqnarray}
   F(k) & = & \langle r=0 | k^{(+)} \rangle   
     =  \langle r=0 | k \rangle + \langle r=0 | G_0(E) T(E)| k \rangle \\
    & = & 1 +
    \frac{\gamma^2 Z(k) \xi(k)}{\displaystyle \frac{k^2}{2m} 
    - E_r - \gamma^2 D(k)}
\label{FCCM}
\end{eqnarray}
with
\begin{eqnarray}
  Z(k) & = & \frac{-2 i m}{k+i\mu} \ .
\end{eqnarray}

\begin{figure}[htb]
\begin{center}
\mbox{\epsfxsize=60mm \epsffile{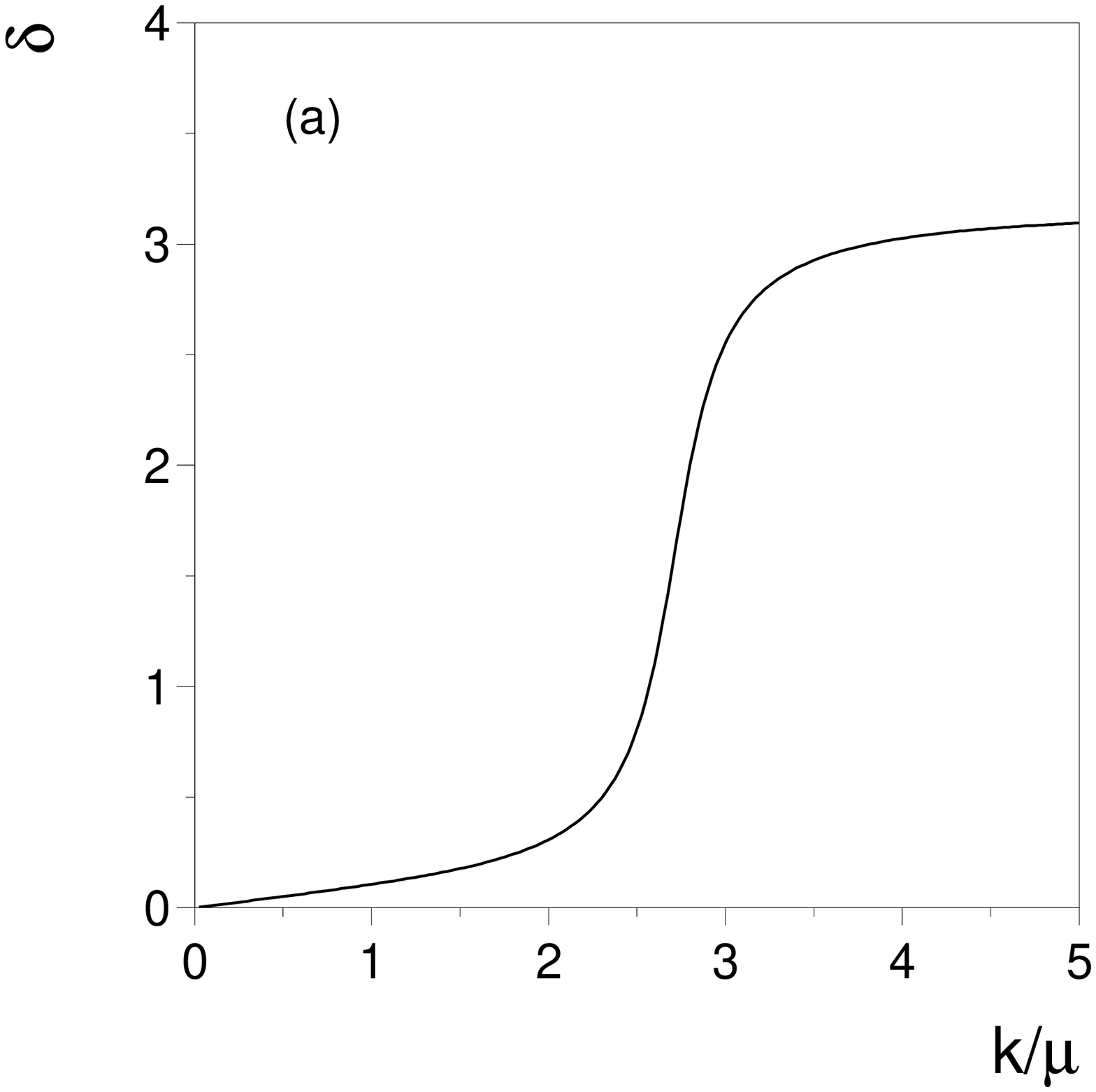}} \hspace*{20mm}
\mbox{\epsfxsize=60mm \epsffile{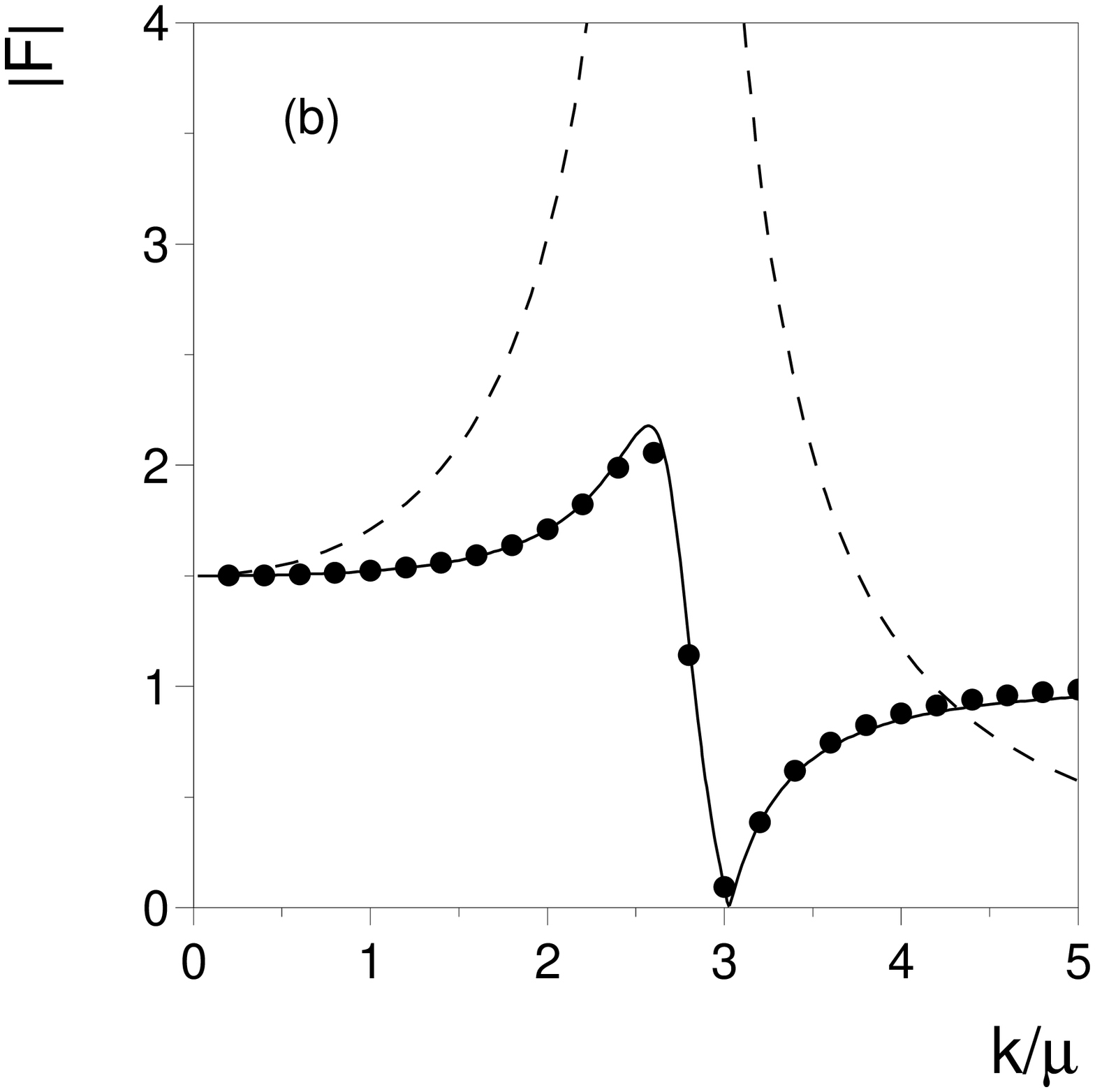}}
\vspace*{-10mm}
\caption{\label{FigCCM}%
The momentum dependence of the scattering phase (a) and the
form-factor (b) for a resonance in the WW model. 
The solid line is the exact solution (\protect\ref{FCCM}), the dashed 
line is the solution of the OM equation without polynomial factor. 
The dots show the approximate solution of the OM equation with the 
factor $(E-E_z)$.}
\end{center}
\end{figure}

  In Fig.~\ref{FigCCM} we show the scattering phase and the form-factor for
the WW model for $\mu=5m$, $\gamma=10\mu^{3/2}$, $E_r=4m$.
  The pole produces a resonance peak in the energy
dependence of the form-factor which  is damped by a nearby zero 
restoring the right limit $F\to 1$ for $E\to\infty$.
The reduction imposed by this zero is enormous. 

Formula (\ref{FCCM}) can be rewritten explicitly showing the
interplay of the pole and the zero:
\begin{eqnarray}
   F(k) & = & \frac{A(E)}{B(E)} \label{FABCCM} \\
   A(E) & = & E - E_r - \frac{\gamma^2 m}{\mu} \frac{1}{(k^2+\mu^2)} \\
   B(E) & = & E - E_r - \frac{\gamma^2 m}{\mu}
                        \frac{(k^2-\mu^2-2i\mu k)}{(k^2+\mu^2)^2} \ .
\end{eqnarray}
 In the limit of weak coupling the resonance in the scattering channel is
directly connected to the bound state in the continuum which has
an energy shift $\Delta E_r$ and a width $\Gamma_r$:
\begin{eqnarray}
   \Delta E_r & = &
          \frac{\gamma^2 m (k^2-\mu^2)}{\mu (k^2+\mu^2)^2}  \\
   \Gamma_r & = &
          \frac{4 \gamma^2 m k}{(k_r^2+\mu^2)^2} 
\end{eqnarray}
where $E_r = k_r^2/2m$.
The form-factor in the vicinity of the resonance has the form
\begin{eqnarray}
   F(k) & = & \frac{E-E_z}{E - (E_r + \Delta E_r - i \Gamma_r /2)}
   \label{FFPZ}
\end{eqnarray}
and the zero $E_z$ is located near the resonance energy $E_r+\Delta E_r$
at
\begin{eqnarray}
   E_z & = & E_r + \Delta E_r + \frac{\mu}{k_r} \Gamma_r
\end{eqnarray}
If $|E_z - E_r - \Delta E_r| > \Gamma_r$, the resonance produces a
pronounced peak followed by a dip in the energy dependence of the
form-factor. In case $|E_z - E_r - \Delta E_r| < \Gamma_r$ the energy
dependence coming from the pole is damped completely by the zero in
the nominator, and only a dip is visible in the form-factor.
Notice that the zero is of dynamical nature and  disappears for 
vanishing channel coupling:  $F\to 1$ as $\gamma\to 0$. 

  Since $A(E)$ is a real symmetric function of momentum $k$, it does  not
contribute to the elastic scattering amplitude. The solution of the OM
equation  without a polynomial factor reflects only  the resonance pole in
formula (\ref{FFPZ}) as shown in Fig.~\ref{FigCCM}, dashed line. 
By including the factor $(E-E_z)$ one gets
\begin{eqnarray}
   F(E) & = & F(0) \frac{(E_z-E)}{E_z}
             \exp \left( \frac{E}{\pi}
             \int_{0}^{\infty} \frac{\delta(E')}{E'(E'-E)}
             dE' \right) 
\end{eqnarray}
which is very close to the exact solution%
\footnote{A careful analysis of the OM equation for the model considered 
shows that there is an extra factor $(k^2+\nu^2)/(k^2+\mu^2)$ resulting 
from the singularities in the upper halfplane of complex momentum $k$:
a pole at $k=i\mu$ and a nearby zero at $k=i\nu$. For our example this factor 
is close to 1 in the region of the resonance.}%
of the WW model. 

  These results characterize a {\it coupled-channel} resonance.  The
scattering phase beyond the resonance does not decrease as it occurs for a
direct channel {\it potential} resonance%
\footnote{In the literature the first category is often
called {\it normal resonance} and the second one {\it molecular}  or
{\it bootstrap resonance}, see {\it e.g.} \cite{AMP87} and references therein.}
where no extra polynomial factor appears in the solution of the OM equation.
For a potential resonance the decrease of the phase for $s\to\infty$ 
guarantees that the asymptotic limit of the form-factor is one. 

  It must be emphasized that in the WW model considered, the resonance-dip
structure occurs only in processes where the particles in the scattering
channel are produced at small distance due to some extraneous interaction
which can be treated perturbatively, so that the momentum dependence of the
production amplitude is entirely determined by the form factor $F(k)$
given by Eq.~(\ref{FCCM}) (this is  relevant for the
$K\to\pi\pi$ decay).
  This situation must be distinguished from a situation where the original
bound state $|b\rangle$ is produced as a resonance with amplitude $C$ and then
decays into the scattering channel. The corresponding amplitude with 
rescattering included is  
\begin{eqnarray}
   T_b(k) & = & C
   \frac{\gamma \xi(k)}{\displaystyle \frac{k^2}{2m} - E_r - \gamma^2 D(k)}
\label{TB}
\end{eqnarray}
which has a purely resonant behavior, there is no nearby zero.
Studying the energy dependence of the data in the vicinity of the resonance
one can determine whether this situation is realized for the process in
question.   

\subsection{Application to the $f_0$ resonance 
and constraint from $pp\to pp\pi\pi$}
\label{Appl_f0}

  To evaluate the role of the $f_0$ resonance for $K\to\pi\pi$ decay we use
the S-matrix in Breit-Wigner form fitted to data \cite{ZB94}, see 
Eq.~(\ref{resf}). 
As we demonstrated in Sec.~\ref{WWM}, the polynomial in the solution of the
OM  equation is expected to  have a zero at $s=s_z$ close to the resonance:
\begin{equation}\label{40}
   P(s) = 1 - \frac{s}{s_z}
\end{equation}
Note that $s_z \to M_r^2$ as $g_1\to 0$.  In order to fix the position of
the zero we use information from a related process and study the effective
mass  distribution ($M=\sqrt{s}$) of pion pairs produced in the reaction 
$pp\to pp\pi\pi$ \cite{AFS},  which can be expressed by \cite{MP84}
\begin{eqnarray}
  \frac{d\sigma}{d M} \sim \frac{(M^2-4 m_{\pi}^2)^{1/2}}{M^3} |F(M^2)|^2
\end{eqnarray}

\begin{figure}[htb]
\begin{center}
\mbox{\epsfxsize=60mm \epsffile{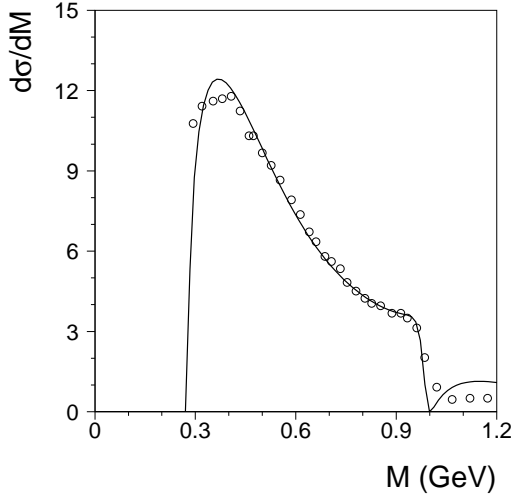}}
\vspace*{-10mm}
\caption{\label{Figprod}
The effective mass distribution of  pion pairs 
in $pp\to pp\pi\pi$ vs. $M=\protect\sqrt{s}$.
The data are from \protect\cite{AFS}.}
\end{center}
\end{figure}

  Including the polynomial~(\ref{40}) into the calculation of the
form-factor $F(s)$ ($f_0$ plus $\rho$ and $f_2$ exchange) we obtain
$s_z=1.0$~GeV$^2$ for the position of the zero, see Fig.~\ref{Figprod}.  The
fit shown for the mass distribution $d\sigma/dM$ also contains a factor
$(1+0.25s)$ in the polynomial and an overall normalization constant. The
position of the zero however, is determined very precisely from nearby data
alone. The corresponding scalar form-factor will be discussed in
sec.~\ref{Sec_Kpipi}.

  Note that the factor containing the zero can be incorporated into a formal
solution of the OM equation, if a physically equivalent discontinuous
scattering phase is introduced 
\begin{equation}
    \bar{\delta}(s) = \delta(s) - \pi \theta(s - s_z) \ .
\label{deltabar}
\end{equation} 
It is interesting to note that the description of the pion pair 
distribution $d\sigma/dM$ in \cite{MP84,AMP87} was seemingly achieved using the 
trivial  polynomial $P(s)=1$. However the elastic phase was calculated from
the expression 
$\Phi=\mbox{\rm Arctan} (\mbox{\rm Im}\;T_{11}/\mbox{\rm Re}\;T_{11})$.
In the presence of
inelasticities the phase of $T_{11}$ is bounded to the interval
$[0,\pi)$  by the requirement of continuity. When $\mbox{\rm Re}\;T_{11}$
changes  from negative to positive due to the sharp resonance rise of
$\delta(s)$  the phase $\Phi$ drops sharply by nearly $\pi$. With this
choice  the Omn\`es  function develops a zero close to the point where
$\delta=\pi$.  While this provides a good description of the data,  the
introduction of the zero in this way appears to be accidental. For instance,
in the model considered  in Sec.~\ref{WWM} there is no connection between
the position of the zero  and the condition $\delta=\pi$. Also, if the
scattering phase $\delta$  reached $\pi$ before the $K\bar{K}$ threshold,
the zero factor would not be  obtained from using the phase prescription
for $\Phi$ quoted above.

\section{Summary of results for the $K\to\pi\pi$ decay and conclusion}
\label{Sec_Kpipi}
   
For the $\Delta I=\frac{1}{2}, \frac{3}{2}$ $K\to 2\pi$ decays, without
long--range final state interactions, the bare weak vertex can be
parametrized as~\cite{Kpipi}, $\Gamma^b_{I=0,2}(s)=C_{I=0,2}(s-m_\pi^2)$
where $s=m_K^2$ for an on-shell kaon.
The quantity $\Gamma^b$ contributes to the polynomial
in the Omn\`es representation of the $K\to\pi\pi$ decay.
  We have prepared the ground for the $I=0$ $S$-wave final state interaction
in the preceeding sections. The solid line in Fig.~\ref{FigFFrhof0} shows
the net result for the model combining the $\rho$ and $f_2$ exchange with
the $f_0$ resonance. The resonance in the form-factor is protected by the
zero at $s_z=1\;$GeV$^2$ as determined from the pion pair production data.
At the kaon mass the $I=0$ enhancement factor is  $F(m_K^2)=1.62$, a result
which is similar to the values obtained in the literature quoted above. 
From $\rho$ exchange alone we obtain $F(m_K^2)=1.38$, $\rho$ and $f_2$ 
give $F(m_K^2)=1.57$ while the enhancement from the $f_0$ resonance alone 
is $F(m_K^2)=1.03$. 
For the complete form-factor the reduction of the $f_0$ contribution induced
by the protective zero is of course crucial. The effects of the zero and the
resonance largely  cancel and only a very small contribution to the
form-factor far away from the pole (zero) survives. For example, at $s=0$
the pion scalar radius is:  
$\langle r_s^2\rangle =0.52\;$fm$^2$ when only considering $\rho$ and $f_2$ 
exchanges. When including the resonance protected by the  zero we have  
$\langle r_s^2\rangle = 0.58\;$fm$^2$. We see that the inclusion of the 
resonance does improve the result on the scalar radius  but avoids too 
large an effect.  Our full result is very close to the value obtained 
in \cite{GM91} where $\langle r_s^2\rangle$ is determined  from chiral 
perturbation theory. Without the zero we would have obtained a rather large 
value $\langle r_s^2\rangle =0.81\;$fm$^2$.

\begin{figure}[htb]
\begin{center}
\mbox{\epsfxsize=60mm \epsffile{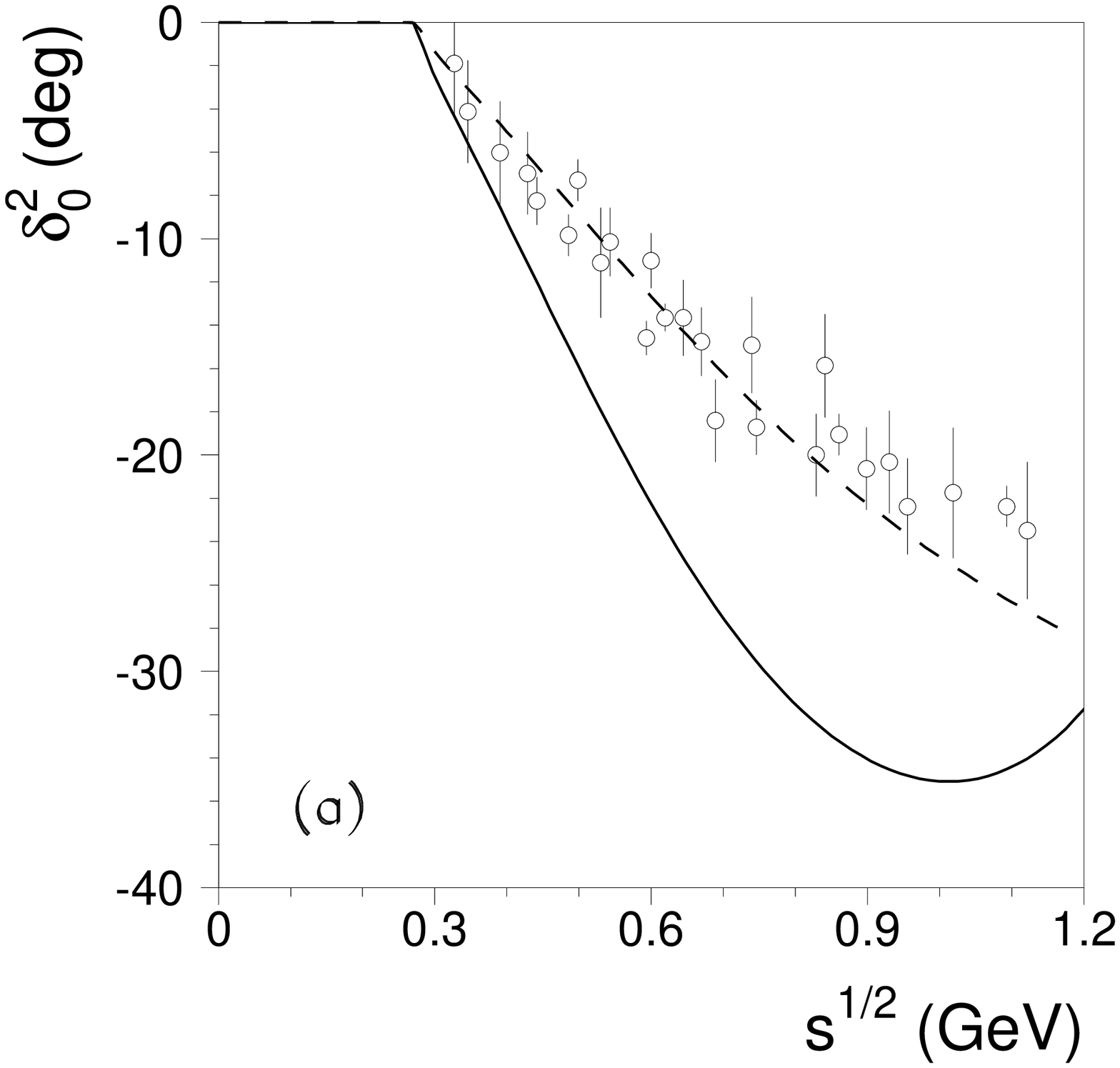}}
\hspace*{10mm}
\mbox{\epsfxsize=60mm \epsffile{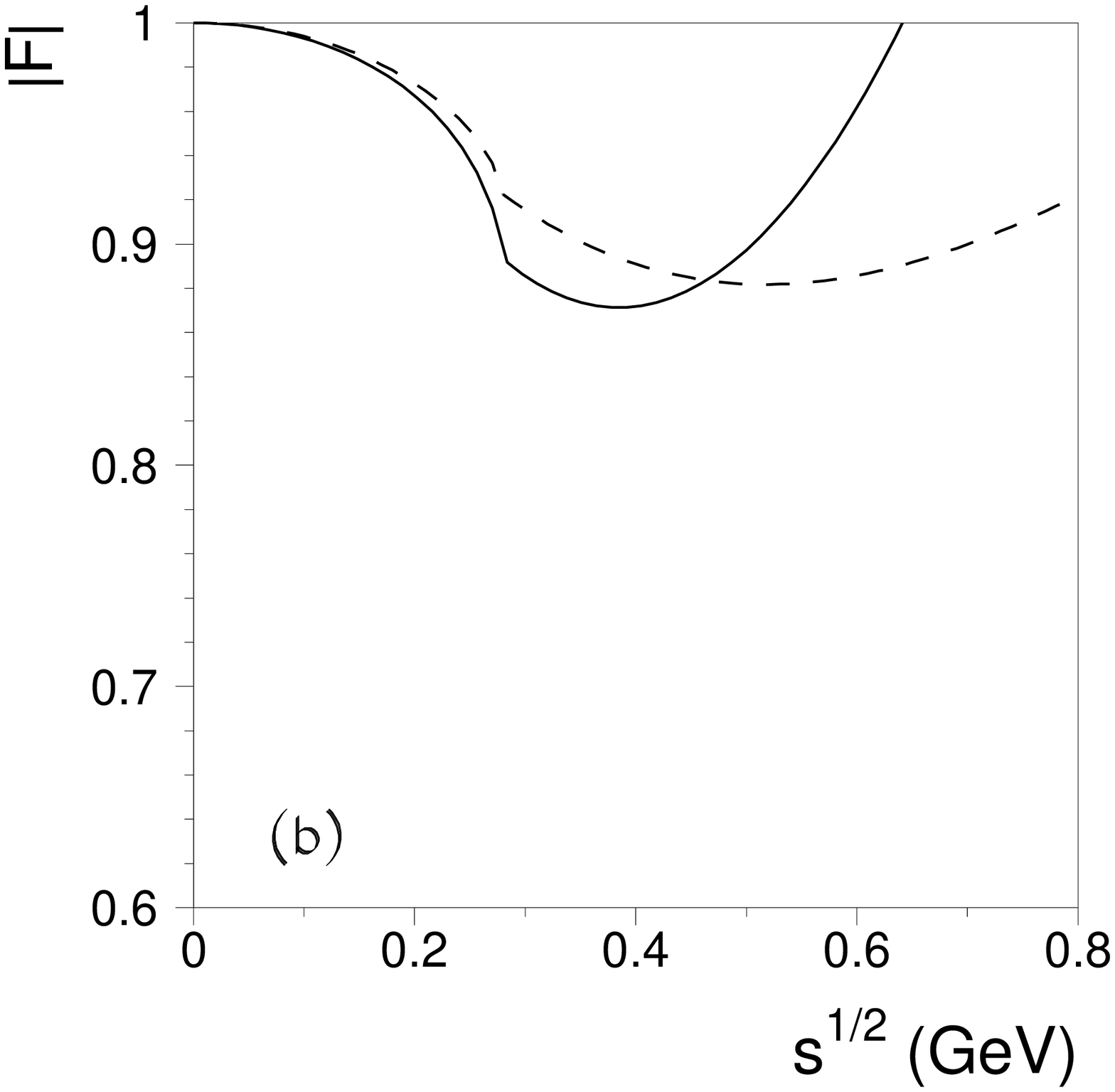}}
\vspace*{-10mm}
\caption{\label{FigI2}
The $\pi\pi$ $I=2$ $S$-wave scattering phase $\delta^{I=2}_0$ 
vs. $\protect\sqrt{s}$ (a) and the form-factor $F^{I=2}(s)$ (b). 
Solid line: $\rho+f_2$ exchange, dashed line:
$\rho$-exchange with vertex form-factor. 
The experimental data are from
\protect\cite{pipiI2J0}.}
\end{center}
\end{figure}

In order to complete the evaluation of the the overall $\Delta I=1/2$ 
enhancement factor the contribution of the $I=2$ channel must be
evaluated as well. Due to the signature of the crossing matrix the
contribution from $\rho$ exchange is repulsive in the $I=2$ channel,
see (\ref{TrhoBA}). On the other hand $f_2$ exchange does not change
sign relative to the $I=0$ channel leading to destructive interference
between $\rho$ and $f_2$ for the isotensor. 
The solid line in Fig.~\ref{FigI2}a
shows the unitarized sum of $\rho$ and $f_2$ exchange. Also shown is
$\rho$ exchange modified by a vertex form factor with monopole range
$\Lambda_{\rho}=1.5\;$GeV (dashed line) which is a good
effective parametrization of the data. The phases at higher energies are not
known, but fortunately the form-factor at $\sqrt{s}=m_K$ is not sensitive to
this region.
   The corresponding form-factor $F^{I=2}(s)$ is shown in Fig.~\ref{FigI2}b.
At the kaon mass we obtain a reduction factor $F^{I=2}(m_K^2)=0.9$
leading to a combined $\Delta I=1/2$ enhancement of
$F^{I=0}(m_K^2)/F^{I=2}(m_K^2)=1.81$ which is satisfactory, 
but slightly less than the value required by the data. 

  We conclude that the $\rho$ and $f_2$ exchange interactions remain
the dominant mechanisms for the FSI enhancement factor in the 
$\Delta I=1/2$ rule in $K\to\pi\pi$. Since $\rho$ exchange generates a 
broad pole in the $I=0$ $S$-wave amplitude \cite{ZB94} one can associate 
this enhancement with a $\sigma$ meson. 
The $f_0$ resonance plays a minor role.  
This is due to the occurrence of a protective zero at 
$s=1$~GeV$^2$ modifying the polynomial in the OM equation. The nature and 
position of this zero has been verified by analyzing pion pair production in 
$pp\to pp\pi\pi$ where the $f_0$ resonance only leads to a small shoulder
in the mass distribution. A simple coupled channel model describes this 
situation very adequately. We expect that this damping mechanism 
will be applicable to many other decay and production reactions in the 
vicinity of a coupled channel resonance.

\section*{Acknowledgements}

The authors thank Bing-Song Zou for useful discussions.


\end{document}